\documentclass[preprint2]{aastex}          
\usepackage{graphicx}

\usepackage{fancyheadings}

\begin{document}
\thispagestyle{fancy}
\def\teff{$T\rm_{eff }$}
\def\kms{$\mathrm {km s}^{-1}$}

\title{Variation rate of sunspot area}


\slugcomment{Proceeding roma}

\lhead{\textit{Proceedings EWASS 2012, Rome\\
The Sun: new tools and ideas in observational solar astrophysics}}
\rhead{\copyright MEM. Soc Astron.\\ Italiana Supplements} 

\author{
Gafeira R.\altaffilmark{1}
\and Fonte C. C. \altaffilmark{2.3}
\and Pais M. A.\altaffilmark{1,4}
\and Fernandes J.\altaffilmark{1,3,5}
          }


\altaffiltext{1}{
Center for Geophysics of the University of Coimbra, Portugal
}
\altaffiltext{2}{
Institute for Systems and Computers Engineering at Coimbra, Portugal
}
\altaffiltext{3}{
Department of Mathematics, University of Coimbra, Portugal
}
\altaffiltext{4}{
Department of Physics, University of Coimbra, Portugal
}
\altaffiltext{5}{
Astronomical Observatory of the University of Coimbra, Portugal
}



\begin{abstract}
The emergence of the magnetic field through the photosphere has multiple manifestations and sunspots are the most prominent examples of this. One of the most relevant sunspot properties, to study both its structure and evolution, is the sunspot area: either total, umbra or penumbra area. Recently \citet{Schlichenmaier2010} studied the evolution of the active region (AR) NOAA 11024 concluding that during the penumbra formation the umbra area remains constant and that the increase of the total sunspot area is caused exclusively by the penumbra growth. In this presentation the Schlichenmaier's conclusion is firstly tested, investigating the evolution of four different ARs. Hundreds of Intensitygram images from the Helioseismic and Magnetic Imager (HMI) images are used, obtained by the Solar Dynamics Observatory, in order to describe the area evolution of the above ARs and estimate the increase and decrease rates for umbra and penumbra areas, separately. A simple magnetohydrodynamic model is then tentatively used in a first approximation to explain the observed results.\\
\\
\end{abstract}



\section{Introduction}

Sunspots are among the most prominent manifestations of interaction between the solar magnetic field and solar plasma. One of the most relevant sunspot properties to study is the structure and evolution of their area, either umbra, penumbra or total area.

\citet{Hathaway2008} computed the daily values of the total area of one active region during 14 days. Recently, \citet{Schlichenmaier2010} studied the evolution of one active region, NOAA 11024. They computed the areas of umbra and penumbra during 6 hours and concluded that during the penumbra formation the umbra area remains constant and the increase of the total sunspot area is caused exclusively by the penumbra growth.

One of the goals of this work is to compute the area evolution of four active regions, during their transit through the visible hemisphere, using images taken at intervals of approximately 10 $\approx$ 20 min. 
We then concentrate on the study of the area evolution after the active regions have reached their maxima, a period that we call decay phase. In order to explain the observations, we propose a simple model whereby a radial flow has a dominant role in the umbra/penumbra evolution, both through advection and stretching of the magnetic field.

\section{Data series and fuzzy area}

\begin{figure}[h]
\centering
\resizebox{7cm}{!}{\includegraphics[clip=true]{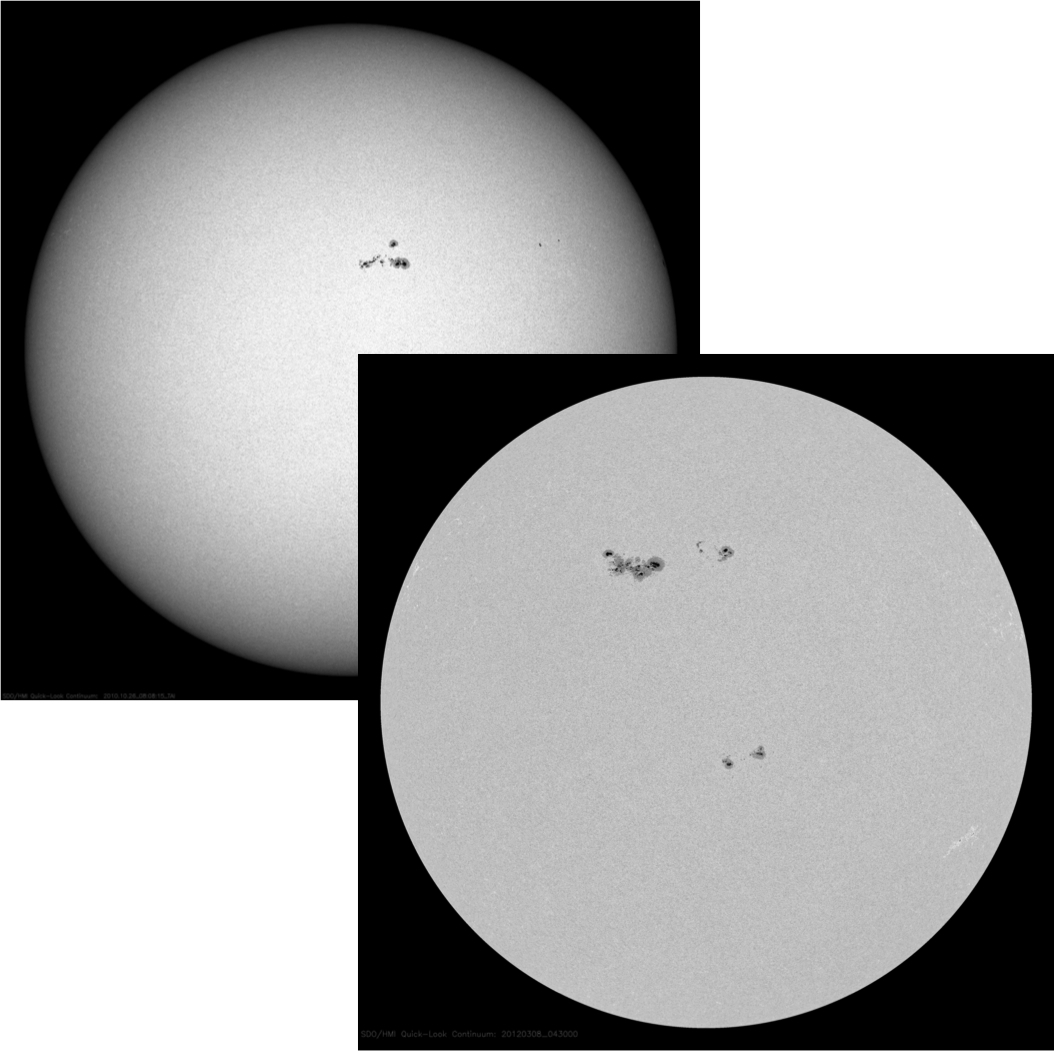}}
\caption{
\footnotesize
Snapshots of the active regions considered. NOAA 11117 in the left panel, NOAA 11428,11429,11430 in the right panel.
"Courtesy of NASA/SDO and the AIA, EVE, and HMI science teams."
}
\label{imexam}
\end{figure}

In this study we use the images from Solar Dynamics Observatory (SDO) with 4096 x 4096 pixels and time steps of approximately 10 to 20 minutes. The observed active regions are NOAA 11117, with observations starting in 26 October 2011, and NOAA 11428, NOAA 11429, and NOAA 11430, with observations starting in 8 March 2012. Figure \ref{imexam} shows examples of the kind of images used for the study of the considered active regions.

\begin{figure}[h]
\centering
\resizebox{7cm}{!}{\includegraphics[clip=true]{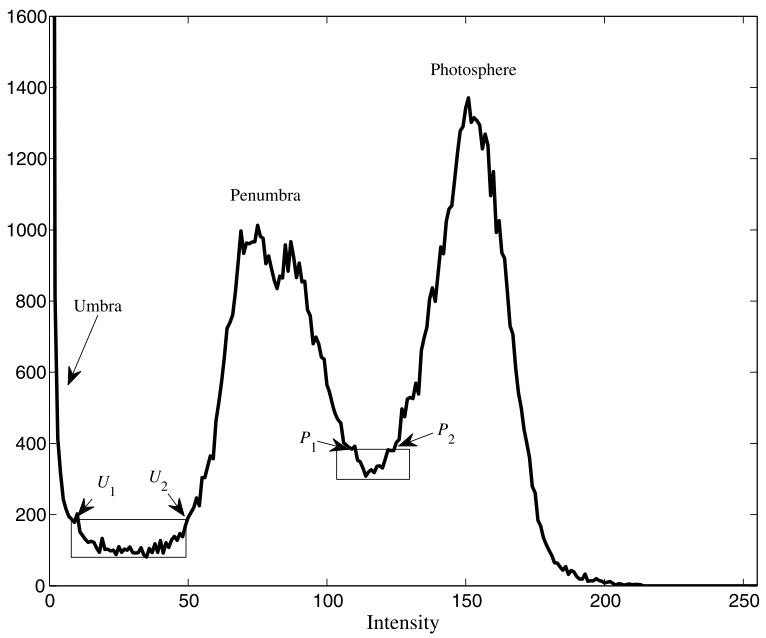}}
\caption{
\footnotesize
Example of the histogram obtained from one smoothed image. 
Values $U_1$ and $U_2$ represent the limits of the region of transition between the umbra and penumbra and values $P_1$ and $P_2$ the limits of the transition between the penumbra and the photosphere.
\citep{Fonte2009}.
}
\label{histfuzzy}
\end{figure}

Due to the difficulty to define the exact limit of the sunspots and in particular to define the limit between umbra and penumbra, we use fuzzy sets to compute the area of the sunspots. Figure \ref{histfuzzy} shows the intensity histogram
obtained from a smoothed image, from which the membership functions to the fuzzy umbra and penumbra may be computed. The degrees of membership of each intensity value to the fuzzy umbra and penumbra are obtained  using values $U_1$, $U_2$, $P_1$ and $P_2$, which delimit respectively the regions of uncertainty corresponding to the separation between the umbra, penumbra and photosphere \citep{Fonte2009}.

\section{Area calculation and simulation}

\begin{figure}[h]
\centering
\resizebox{7cm}{!}{\includegraphics[clip=true]{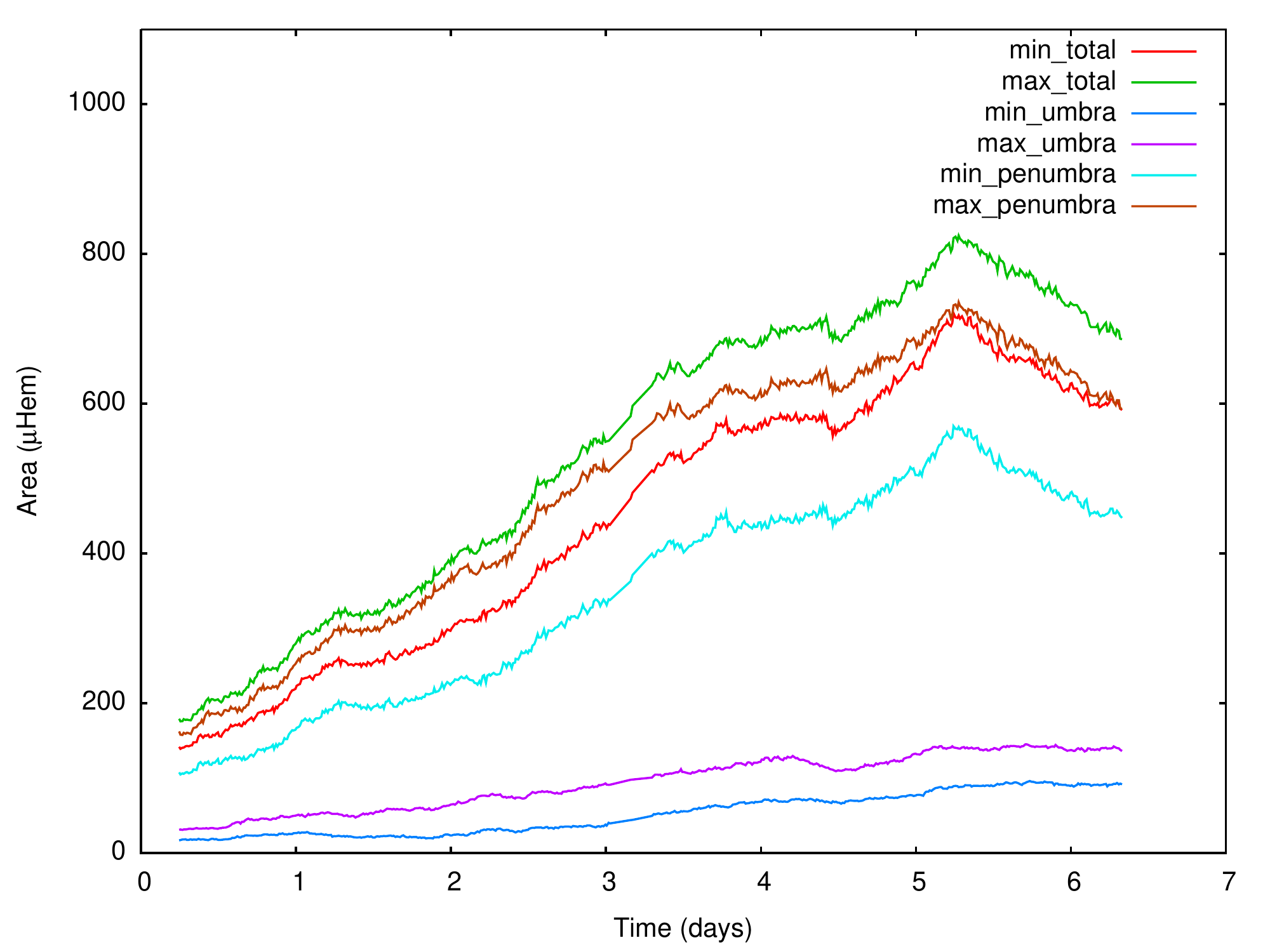}}
\caption{
\footnotesize
Area evolution of the active region NOAA 11117.
}
\label{11117}
\end{figure}

\begin{figure}[h]
\centering
\resizebox{7cm}{!}{\includegraphics[clip=true]{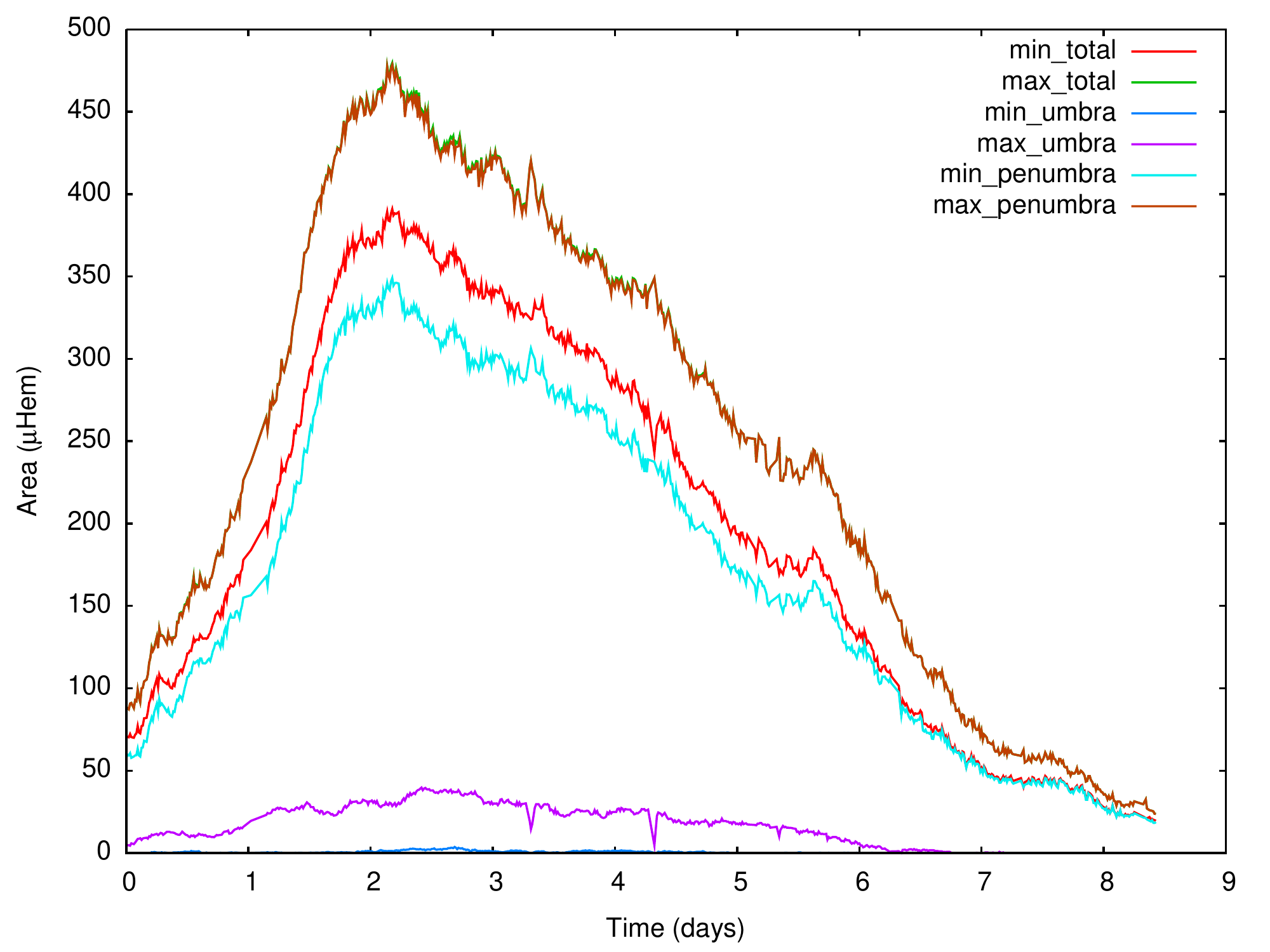}}
\caption{
\footnotesize
Area evolution of the active region NOAA 11428.
}
\label{11428}
\end{figure}

\begin{figure}[h]
\centering
\resizebox{7cm}{!}{\includegraphics[clip=true]{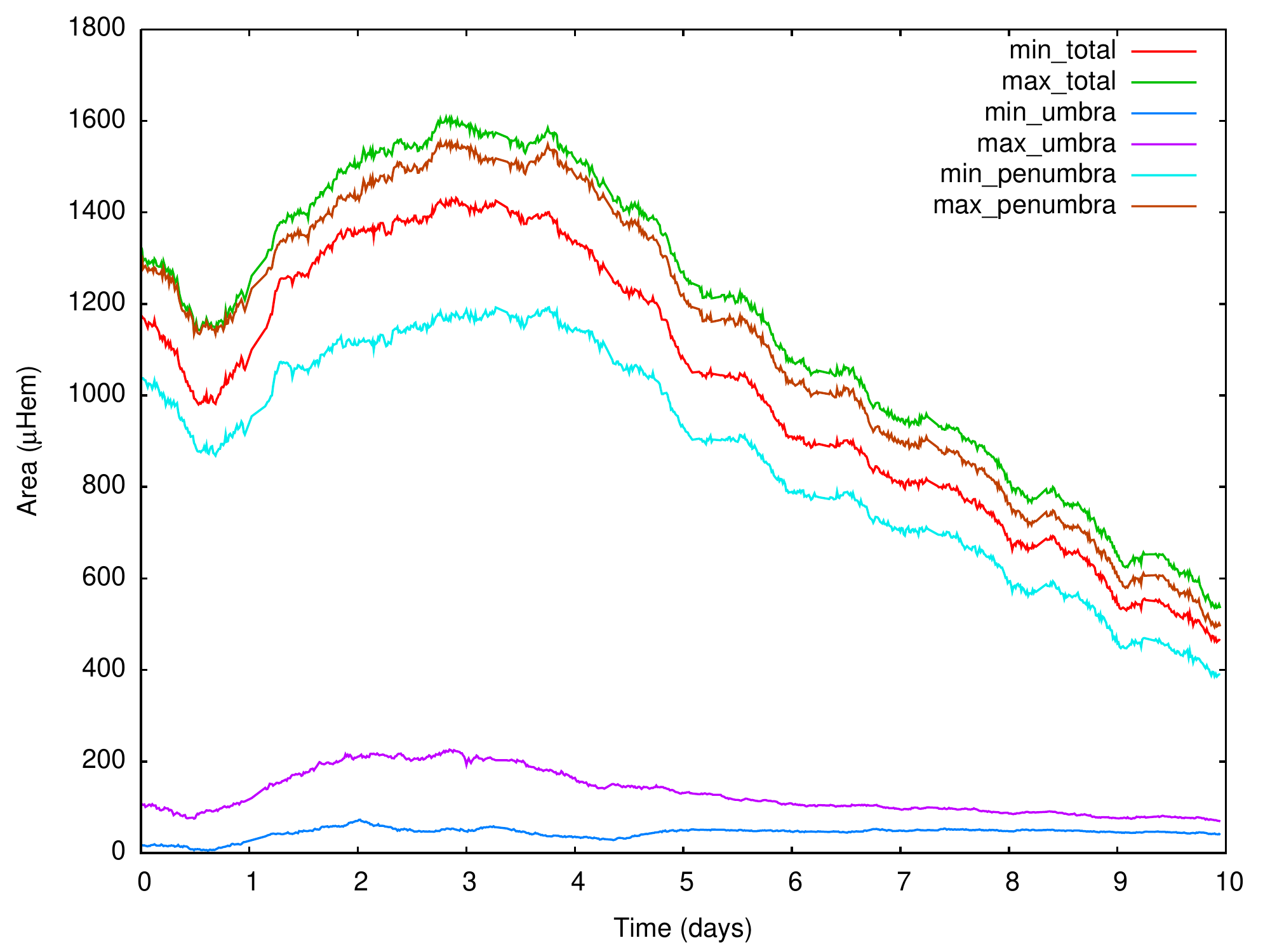}}
\caption{
\footnotesize
Area evolution of the active region NOAA 11429.
}
\label{11429}
\end{figure}

\begin{figure}[h]
\centering
\resizebox{7cm}{!}{\includegraphics[clip=true]{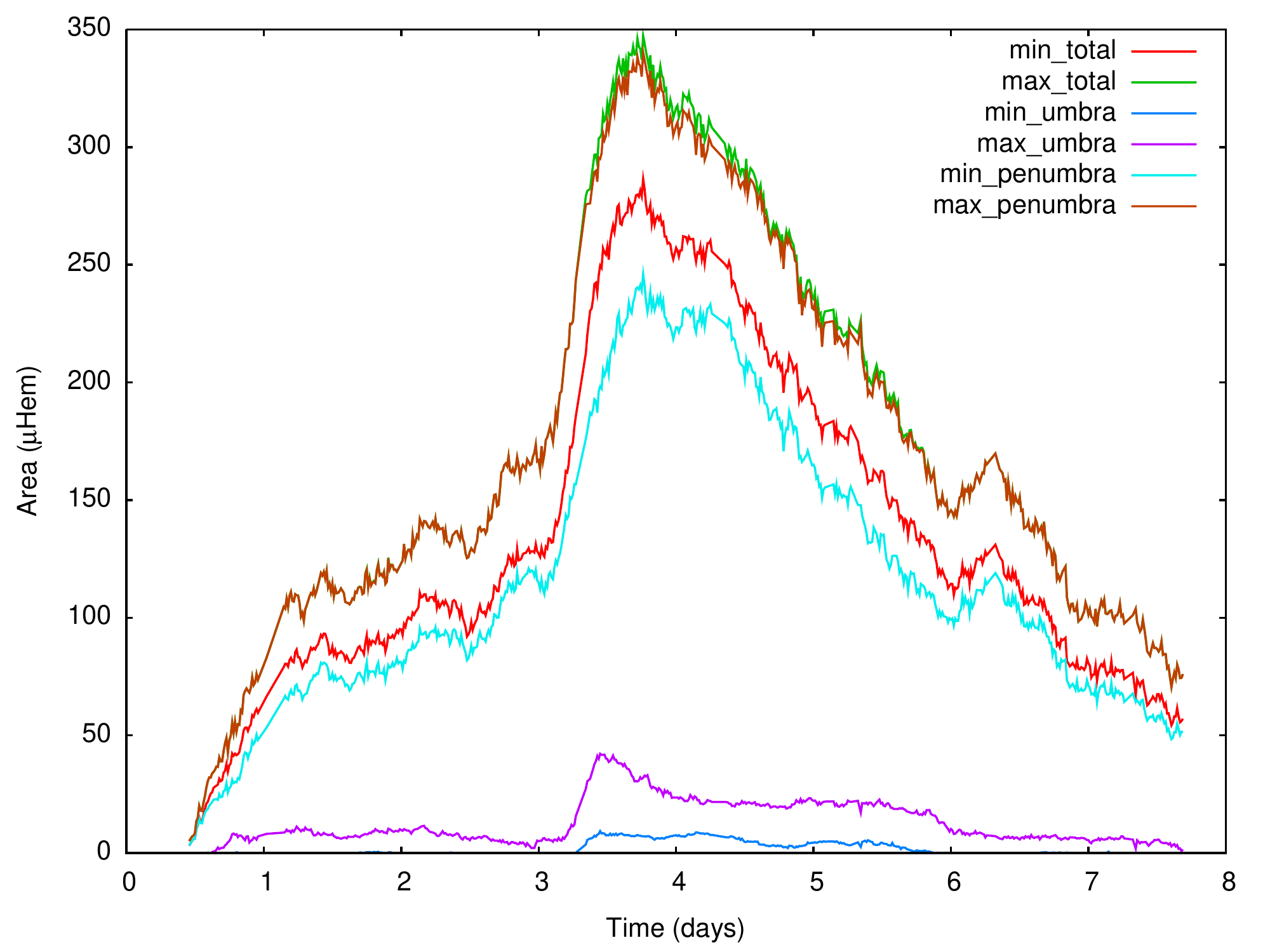}}
\caption{
\footnotesize
Area evolution of the active region NOAA 11430.
}
\label{11430}
\end{figure}

The areas of the regions corresponding to the fuzzy umbra, penumbra and the total sunspot of the four active regions indicated below are computed using a fuzzy approach, which provides information on the uncertainty of the obtained values \citep{Fonte2009, Fonte03areasof}. 
The obtained results are shown in figures \ref{11117}, \ref{11428}, \ref{11429} and \ref{11430}, where the variation of the minimum and maximum possible values for the area of the total sunspot (respectively the red and green lines), the umbra (respectively the dark blue and purple lines) and the penumbra (respectively the light blue and brown lines) are represented.

Using the information extracted from the evolution of the active region area, we compute the growth rate for the maximum umbra and the maximum penumbra areas and the decay rate for the maximum area of the total active region using a linear fit. The results are described in table \ref{lisfit}. 

\begin{table}[h]
\caption{Results obtained for the growth rate, decay rate and maximum area for each active region.}
\label{lisfit}
\begin{center}
\begin{tabular}{cccc}
\hline
\\

NOAA &Max area  & Umbra &Total area\\
&& growth rate &decay rate\\
&$(\mu Hem)$&$(\mu Hem/day)$&$(\mu Hem/day)$\\
\hline
\\
11117  & $800$ & $21$ & $-127$\\
11428  & $480$ & $12$ & $-85$\\
11429  & $1600$ & $40$ & $-142$\\
11430 & $350$ & $6$ & $-84$\\
\\
\hline
\end{tabular}
\end{center}
\end{table}

In a first approach to interpret the obtained results, a simple diffusion-advection numerical model using the MHD formalism was used. In this model we simulate a circular formed sunspot and extract the area evolution for the decay phase.

We assume a circular sunspot with only radial velocity of the form $u_r =u_0 r/r_0$, where $r$ is the radial coordinate and $r_0$ is the initial radius of the sunspot. The system of equations to solve in cylindrical coordinates $(r,\theta,z)$ assuming $\partial/\partial z=\partial /\partial \theta=0$ is represented by equations 1 and 2,

\begin{equation}
\label{difz}
\frac{\partial B_z}{\partial t} = -u_r \frac{\partial B_z}{\partial r} + \eta \left(\frac{\partial^2 B_z}{\partial r^2}+\frac{1}{r}\frac{\partial B_z}{\partial r}\right)
\end{equation}

\begin{eqnarray}
\label{difr}
\frac{\partial B_r}{\partial t} &=&B_r \frac{\partial u_r }{\partial r} -u_r \frac{\partial B_r}{\partial r} \nonumber\\
&& + \eta \left(\frac{\partial^2 B_r}{\partial r^2}+\frac{1}{r}\frac{\partial B_r}{\partial r} -\frac{B_r}{r^2}  \right)
\end{eqnarray}
where $\eta$ is the magnetic diffusivity and has the value of 0.2 $km^2  s^{-1}$ \citep{Chae2008}.

The initial field considered is constructed based on the observation of NOAA 10933 from \citet{Borrero2011}, where we adjust the data to a linear combination of Bessel functions of the first kind. The expressions that we obtain are represented in equations \ref{campz} and \ref{campr},

\begin{eqnarray}
\label{campz}
B_z &=& 700J_0 \left(  \frac{9r}{1.7r_0} +   \frac{4r}{1.7r_0}   \right. \nonumber\\
&&  \left. +  \frac{5r}{1.7r_0}  +   \frac{2.2r}{1.7r_0} -0.1390 \right)
\end{eqnarray}

\begin{equation}
\label{campr}
B_r = 900J_1\left(\frac{9r}{1.7r_0} +  \frac{4r}{1.7r_0} + \frac{5r}{1.7r_0} \right)
\end{equation}
where $J_0$ and $J_1$ are Bessel functions of the first kind of order zero and one respectively.

\begin{figure}[h]
\centering
\resizebox{7cm}{!}{\includegraphics[clip=true]{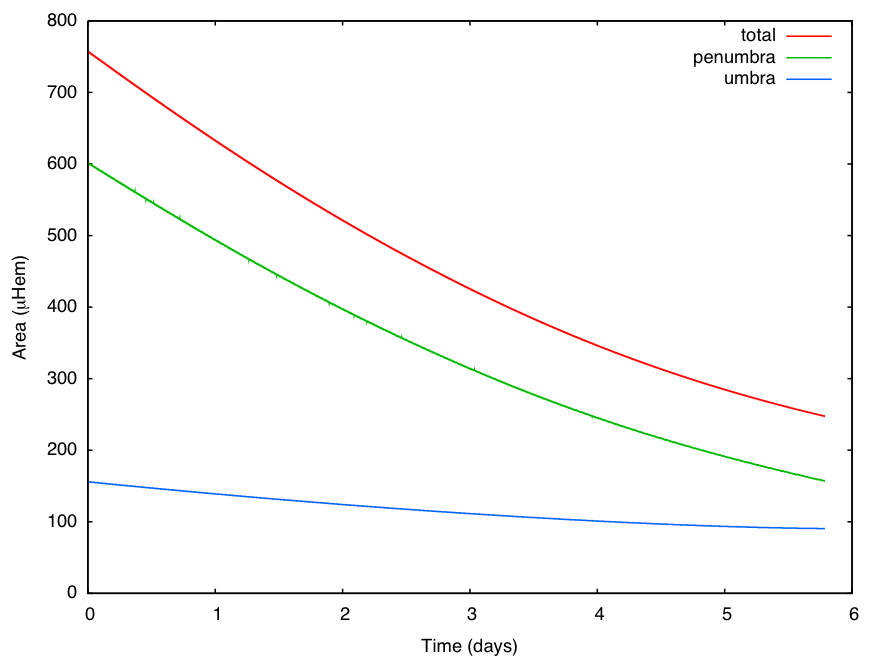}}
\caption{
\footnotesize
Example of the results obtained using the model.
}
\label{simgraf}
\end{figure}

The results obtained in the simulation using $u_0$ equal to -0.035 $Km ~ s^{-1}$ are represented in table \ref{lissim} and in figure \ref{simgraf}. The value for $u_0$ is adjusted in order to produce the best agreement between the observation and the simulation result.

\begin{table}[h]
\caption{Results obtained using the model compared to the results obtained by fuzzy method using the observations.}
\label{lissim}
\begin{center}
\begin{tabular}{ccc}
\hline
\\

NOAA &Simulated  &Total area\\
&decay rate&decay rate\\
&$(\mu Hem/day)$&$(\mu Hem/day)$\\
\hline
\\
11117  & $-94$ & $-127$\\
11428  & $-61$ & $-85$\\
11429  & $-143$ & $-142$\\
11430 & $-51$ & $-84$\\
\\
\hline
\end{tabular}
\end{center}
\end{table}

\section{Conclusions}

Our results seem to be consistent with \citet{Schlichenmaier2010} results, though not conclusive yet. Although the umbra growth rate is much weaker than the penumbra's, there is some evolution that can be noticed.

The numerical model tested here indicates that the decay rate of the sunspots could be explained by advection and stretching of the magnetic field due to a negative radial velocity field.


\bibliographystyle{aa}

\begin{thebibliography}{}

\bibitem[Borrero et al. (2011)]{Borrero2011} Borrero, J., Ichimoto, K., \ 2011, Living Rev. Solar Phys., 8, 4

\bibitem[Chae et al. (2008)]{Chae2008}
Chae, J., Litvinenko, Y., Sakurai, T.,\ 2008 ,\aj, 683(2), 1153

\bibitem[Fonte \& Lodwick (2004)]{Fonte03areasof}
Fonte, C.,  \& Lodwick, W. \ 2004, International Journal of Geographical Information Science, Int. J. Geogr. Inf. Sci. ,18, 127

\bibitem[Fonte \& Fernandes (2009)]{Fonte2009} Fonte, C., Fernandes, J., \ 2009, Solar Physics, 260, 21

\bibitem[Hathaway \& Choudhary (2008)]{Hathaway2008}
Hathaway, D.,  Choudhary, D.\ 2008, Solar Physics, 250(2), 269

\bibitem[Schlichenmaier et al. (2010)]{Schlichenmaier2010} 
Schlichenmaier, R.,  Gonzalez, N., Rezaei, R., Waldmann, T.,\ 2010, Astronomische Nachrichten, 331(6), 563





\end{thebibliography}

\end{document}